\documentclass{hotnets19}
\usepackage{times}  
\usepackage{hyperref}
\usepackage{xspace}
\usepackage{balance}
\usepackage{xcolor}
\hypersetup{pdfstartview=FitH,pdfpagelayout=SinglePage}

\usepackage{authblk}
\usepackage{url}

\usepackage{breakurl}

\newcommand{\name}{O-LAD\xspace}
\begin{document}
\balance
% \conferenceinfo{HotNets 2017} {}
% \CopyrightYear{2017}
% \crdata{X}
% \date{}

%%%%%%%%%%%% THIS IS WHERE WE PUT IN THE TITLE AND AUTHORS %%%%%%%%%%%%

\newcommand{\vyas}[1]{~[\textbf{\textcolor{blue}{VS: #1}}]~}

\newcommand{\RD}[1]{~[\textbf{\textcolor{red}{Ram: #1}}]~}

\newcommand{\Rone}[1]{~[\textbf{\textcolor{red}{Reviewer 1: #1}}]~}
\newcommand{\Rtwo}[1]{~[\textbf{\textcolor{red}{Reviewer 2: #1}}]~}
\newcommand{\Rthree}[1]{~[\textbf{\textcolor{red}{Reviewer 3: #1}}]~}
\newcommand{\Rfour}[1]{~[\textbf{\textcolor{red}{Reviewer 4: #1}}]~}

\pagenumbering{arabic}

\title{Fighting Fire with Light: A Case for Defending DDoS Attacks Using the Optical Layer}

% \author{Matthew Hall$^1$, Ramakrishnan Durairajan$^1$, Vyas Sekar$^2$\\
% University of Oregon$^1$, CMU$^2$
% }
\author{
        Matthew Hall$^{\dag}$,
        Ramakrishnan Durairajan$^{\dag}$, 
        Vyas Sekar$^{\ddagger}$\\
        $^{\dag}$ University of Oregon, 
        $^{\ddagger}$ Carnegie Mellon University
    % \\
        % \{mhall,ram\}@cs.uoregon.edu,
        % vsekar@andrew.cmu.edu
}

% \author{
%     \IEEEauthorblockN{
%         Matthew Hall\IEEEauthorrefmark{1},
%         Ramakrishnan Durairajan\IEEEauthorrefmark{2}, 
%         Vyas Sekar\IEEEauthorrefmark{3}
%     }\\
%     \IEEEauthorblockA{
%         University of Oregon\IEEEauthorrefmark{1}\IEEEauthorrefmark{2}
%         CMU\IEEEauthorrefmark{3}
%     \\
%         \IEEEauthorrefmark{1}mhall@cs.uoregon.edu,
%         \IEEEauthorrefmark{2}ram@cs.uoregon.edu,
%         \IEEEauthorrefmark{3}vsekar@andrew.cmu.edu,
%     }
% }

\maketitle

%%%%%%%%%%%%%  ABSTRACT GOES HERE %%%%%%%%%%%%%%

\begin{abstract}

The DDoS attack landscape is growing at an unprecedented pace. Inspired by the recent advances in optical networking, we make a case for optical layer-aware DDoS defense (\name) in this paper. Our approach leverages the optical layer to isolate attack traffic rapidly via dynamic reconfiguration of (backup) wavelengths using ROADMs---bridging the gap between (a) evolution of the DDoS attack landscape and (b) innovations in the optical layer (e.g., reconfigurable optics). We show that the physical separation of traffic profiles allows finer-grained handling of suspicious flows and offers better performance for benign traffic in the face of an attack. We present preliminary results modeling throughput and latency for legitimate flows while scaling the strength of attacks. We also identify a number of open problems for the security, optical, and systems communities: modeling diverse DDoS attacks (e.g., fixed vs. variable rate, detectable vs. undetectable), building a full-fledged defense system with optical advancements (e.g., OpenConfig), and optical layer-aware defenses for a broader class of attacks (e.g., network reconnaissance).

\end{abstract}
\sloppypar

\section{Introduction}

Distributed denial-of-service (DDoS) attacks are on the rise~\cite{rise, mirai_retrospect, mirai_cost, github_attack}. The immense attack volumes that saturate the infrastructure (e.g., transit link flooding), the attack heterogeneity (e.g., distinguishable vs. indistinguishable, direct vs. indirect, etc.), and the low costs to facilitate large-scale attacks (e.g., attacker-defender cost asymmetry) make DDoS {\em the} most important cybersecurity issue faced by today's enterprises. 

% DDoS attacks are one of the greatest cybersecurity threats on the internet today, with the potential to cause millions of dollars in damages. This is evident by recent attacks such as the Mirai attack on DYN in 2016~\cite{mirai_retrospect, mirai_cost}, and the historical growth in strength of attacks, leading to the terabyte attack on GitHub in 2018~\cite{github_attack}. 

Great progress has been made in devising DDoS mitigation strategies. Advances in this front range from well-known techniques such as scrubbing~\cite{Scrubbing_Akamai, Scrubbing_AWS, Scrubbing_CenturyLink, Scrubbing_Cloudflare, kumar2012mitigating} and filtering~\cite{baker2004ingress, ferguson2000network, jin2003hop, yaar2006stackpi} to the recent {\em routing around congestion} (RAC) technique~\cite{smith2018routing}. Despite these advances, the attack landscape is continuously evolving and, as a result, creating a ``silver bullet" solution to tackle DDoS has remained beyond our grasp. For example, Tran et al.~\cite{tran2019feasibility} recently showed that RAC-based DDoS defense is infeasible and unusable in an inter-domain setting. This mandates a rethinking of nature of the DDoS attacks and calls for new defense strategies.

Meanwhile, the optical community has advanced to the point where scaling from 100G to 400G---programmatically and on demand---is possible today~\cite{history}. %As another example, the throughput gain estimations for variable-rate (adaptive) transceivers~\cite{ives2015routing, ives2017throughput, filer2016elastic} are known yet not commonly deployed. 
As another example, the improvements for amplifier modeling~\cite{ikhsan2018performance} and tuning~\cite{xiang2018joint} point towards a rapidly reconfigurable long-haul backbone in the near future. Lastly, wavelength selective switches~\cite{strasser2010wavelength} and reconfigurable add-drop multiplexers (ROADMs) allow wavelengths to change and enable traffic re-routing on the order of microseconds~\cite{porter2013integrating}. While the optical technologies have proven to be of great utility within networking efforts~\cite{porter2013integrating, singla2010proteus, durairajan2018greyfiber}, to the best of our knowledge it has not received enough attention for cybersecurity issues, in general, and DDoS defenses, in particular.

We, therefore, believe it is time to introduce ``optical layer-awareness" to effectively combat DDoS attacks. In this paper, we make a case for an optical layer-aware DDoS defense (\name). The core of \name is based on two key properties of reconfigurable optics: {\em P1} - physical separation of traffic within a shared or congested link~\cite{awduche2001multiprotocol, lee2015routing, zervos2019new}, and {\em P2} - opportunistic reconfigurability of wavelengths~\cite{porter2013integrating, strasser2010wavelength}. By leveraging these two properties---contrary to Tran et al.~\cite{tran2019feasibility}---we posit that RAC-style DDoS defense, while infeasible in an inter-domain setting, is {\em indeed} feasible in an intra-domain, enterprise setting. 

In our preliminary evaluation, we present models for throughput and latency with \name. Our models separate traffic into two groups, suspicious and trusted (utilizing P1), and reconfigures the network topology (with P2). We demonstrate the efficacy of the models on two types of DDoS attacks: direct and indirect. For direct attacks, we reroute the suspicious traffic to a scrubber and send trusted traffic directly to the destination---improving throughput by 25 to 51\%, while reducing latency by 33 to 65\%. %The precise benefit depends on the proportion of trusted good traffic in the network during the attack. 
Similarly, we show how to apply the two proprieties to reduce attacker detection time by 5 to 10$\times$ for indirect (link-flooding) attacks.

%Specifically, in our preliminary evaluation, we show that \name increases throughput for benign traffic proportionally to the fraction of all good traffic that is trusted. For example, when there is 100 Gbps of attack traffic against a victim with only 10 Gbps of downstream bandwidth who employs a scrubbing service that guarantees 40 Gbps of traffic filtering, throughput can increase from 37\% when no traffic is trusted and all traffic is filtered at the scrubber, to 62\% or 88\% when 10\% or 80\% of traffic is trusted, respectively.  Under similar conditions, latency for benign traffic decreases as more trusted traffic is routed directly to the destination, and all suspicious traffic is mitigated with the traditional state of the art defenses. For example, if we assume that latency from the network's ingress router where suspicious traffic originates, and egress where the victim resides, is 100 ms, and 100 ms of additional delay is incurred by the scrubber, latency can fall from 535 ms to 361 ms if 40\% of traffic is trusted, or to 187 ms if 80\% is trusted.

While our preliminary results demonstrate the feasibility of \name, a number of grand challenges remain at the intersection of networking, security and optical communities. First, apart from the direct vs. indirect DDoS, challenges lie ahead in modeling and evaluating O-LAD gains for other types of DDoS attacks including (in)distinguishable, volume-based, and protocol-conforming attacks. Second, in addition to DDoS attacks, we posit that O-LAD is applicable for a broader class of cybersecurity issues such as network reconnaissance.

% Distributed denial of service (DDoS) attacks are a common and persistent threat on the Internet. Since their emergence in 1998~\cite{DDoS_Timeline}, no silver-bullet solution has solved the problem that these attacks pose. Today, there are a variety of different DDoS attacks. Some attacks target a single enterprise server or network directly, and others target the victim indirectly via coordinated link-flooding attacks (e.g., Crossfire~\cite{kang2013crossfire}). Attackers can be distinguishable or indistinguishable depending on the scale and type of the attack, and the defense strategy of the victim. For instance, recent work has shown how to identify DDoS attackers in a coordinated link-flooding attack via bandwidth scaling at congested links~\cite{kang2016spiffy}. However, if the attackers have sufficient attack strength to mimic the rate-increase behaviors of legitimate senders (e.g., a state-sponsored attack could conceivably have arbitrarily high attack strength~\cite{History_of_State_Sponsored_Hacking, Gallagher_DDoS_US_NK, Sanger_NK_Cyberpower}), then that attacker may not be identified with the most sophisticated defense strategies. The best DDoS defense solutions of our day critically lack a robust tool to fight the various types of attacks that occur daily in the wild. 

\section{Why Consider Optics?}

Introducing optical layer awareness to the higher layers of the protocol stack has a number of key benefits for DDoS defenses, in particular, and networked systems, in general.

{\bf Optical layer can enable new, more powerful DDoS defenses.} There are two fundamental properties of optics that we can leverage to defend DDoS attacks effectively. First, we can {\em physically separate} traffic (P1), e.g., on different colored lambdas of a shared or congested link. This separation enables us to use optical circuit switches to re-route suspicious traffic to edge-defense appliances such as scrubbers. Subsequently, we can scrub the suspicious traffic while routing the trusted traffic directly to the intended destination, similar to RAC~\cite{smith2018routing}. This way, the trusted traffic benefits from lower latency. Suspicious traffic, if it is determined to be benign, benefits from less congestion (as less traffic is moving through the scrubber). A detailed analysis of this scenario is presented in \S~\ref{sec:througput_analysis}.
    
Second is the {\em opportunistic reconfigurability} of the optical layer using the available backup wavelengths in a network (P2). Wavelength-selective switches (WSS) or reconfigurable add-drop multiplexers (ROADMs) allow wavelengths to change and re-route traffic on the order of microseconds~\cite{porter2013integrating}. We envision this capability enabling the next generation of highly dynamic networks. In these future generation networks, when attackers target a link with a link flooding attack (e.g.,~\cite{kang2013crossfire}), a network controller can quickly identify underutilized, backup, or low-priority wavelengths adjacent to the flooded link to allocate new capacity for trusted traffic on the targeted link. This strategy has the potential to increase the attack cost radically and protect the link (and thus the legitimate flows using it) from congestion.

{\bf Optical layer awareness can make networked systems more efficient.} First, exposing the optical layer to the networked systems has been shown to help operators prevent link failures in backbones~\cite{singh2018radwan} and design better traffic engineering solutions~\cite{ghobadi2016optical}. Second, free-space optics solutions have been shown to reduce latency for intra-datacenter transfers~\cite{ ghobadi2016projector,hamedazimi2013patch}. Finally, recent dynamic capacity planning efforts demonstrate the benefits of reconfigurable optical networks vs. traditional, statically-provisioned networks~\cite{durairajan2018greyfiber}.

{\bf Optical and networking layers are disconnected and their co-optimization is largely unexplored.}
Networked systems depend on the optical layer to support bandwidth-intensive applications and scenarios (e.g., machine learning, volumetric DDoS attacks). These systems generally perform critical functions {\em only} at the network layer e.g., to minimize latency across the network~\cite{jalaparti2016dynamic,liu2014traffic} or to reduce the impact of severe attacks outages ~\cite{kang2016spiffy}. %While these solutions are as compelling as ever, they are oblivious to the underlying optical layer. %layer-3 protocols enable such essential functions with little to no knowledge of the physical layer underneath. 
While the recent studies have looked into joint optimization between the optical and network layers~\cite{durairajan2018greyfiber, jin2016optimizing,singh2018radwan}, the area is still largely unexplored and, in turn, calls for cross-layer solutions. Industry trends indicate a growing interest in optics in network management~\cite{filer2016elastic, klonidis2015spectrally}. We suspect that optical layer management via higher level control is inevitable, and we look towards innovation in this front as an enabler for \name.

\subsection{Open Challenges}
\label{sec:challenges}
Despite these benefits, leveraging optical layer to defend DDoS has its own challenges, which we outline below.

{\bf Optical layer lacks robust APIs.} Networked systems will require a new interface, and modes of cross-layer communication to enable the next generation of networked services. Industry efforts such as OpenConfig~\cite{openconfig} are working to bridge this gap by providing vendor-neutral APIs for network management. However, without sufficient attention from the networked systems and optical communities, these efforts could potentially stagnate. Lack of APIs in the optical layer is a significant open problem which we do not look into in this paper. In this work, we assume that programmable topologies with vendor-neutral APIs exist and explore ways to leverage that ability to fight DDoS attacks.

\section{A Case for Optical Layer-Aware DDoS Defense (\name)}

% \vyas{this section does not add much value. maybe merge sec 3 and sec 4. and make current 3 the vision and the current 4 the analytical basis? or just keep the current title of sec 3 and make sec 4 as the analysis to back it up?} 

{\bf Overview.} In this work, we present a case for Optical Layer-Aware DDoS Defense (\name). % We believe this awareness is essential for combating the ever-growing strength of DDoS attacks (e.g., by opportunistically leveraging reconfigurable low-priority wavelength or by lighting up backup wavelength). 
The key insight in \name is to transmit the trusted traffic over a physically distinct wavelength from suspicious and malicious traffic by opportunistically reconfiguring low-priority wavelengths or backup wavelengths, leading to performance benefits for a victim (i.e., higher throughput and lower latency). 

{\bf Feasibility.} \name is feasible in today's enterprise networks. In what follows, we identify two sources for wavelengths that can be leveraged to implement \name but leave implementation details to future work. First, enterprise networks commonly deploy backup wavelengths, e.g., for fault tolerance and fast fail-over capabilities~\cite{mahimkar2011bandwidth, von2015coronet}. Since these wavelengths are already designated to mitigate link outages, we see them as suitable sources for additional capacity during an attack. Second, wavelengths carrying low-priority traffic can be re-allocated dynamically, away from their path and onto the attack path. Sacrificing low priority traffic during outages is a common theme in traffic engineering (TE)~\cite{hong2013achieving, jain2013b4, kandula2014calendaring} that we appropriate at the optical layer, thus allowing more capacity for trusted traffic during an attack.

% \section{DDoS Attack Models}
\subsection{Analysis of \name}
\label{sec:models}

% \vyas{this section title is vague. can we make such that it actually ties to the message of the paper? analysis and early promise or something like that? } 

%In this section, 
% We begin by formalizing our definition for the enterprise network that can benefit from \name, and then describe DDoS attack models for direct and indirect attacks. Along with these models, we describe state-of-the-art solutions to mitigating the attacks and discuss the efficacy of \name.

% \subsection{Definitions}
% \label{subsec:definitions}
% We formalize the terms necessary
{\bf Definitions.} To quantify the benefits of \name
% % First, defining the graph structure of the AS, and then describing the classes of traffic that we consider. 
we model the enterprise, where \name will be deployed, as a multi-graph $G = (V, E)$. In $G$, $V$ is a set of routers and switches. $E$ is a multi-set of ordered pairs, i.e., $E = \{(e,c)\}$. Let $e$ be an un-ordered pair of switches, $e = \{x, y\} : x,y \in V$, which represents a link (wavelength) between switches, and $c \in C$ be the capacity of the wavelength. $C = \{1, 10, 25, 40, 80, 100\}$ is the set of capacities available for wavelengths with today's commodity transceivers~\cite{fsdotcom}.

\begin{figure}[htb!]
    \center
    \includegraphics[width=0.5\columnwidth]{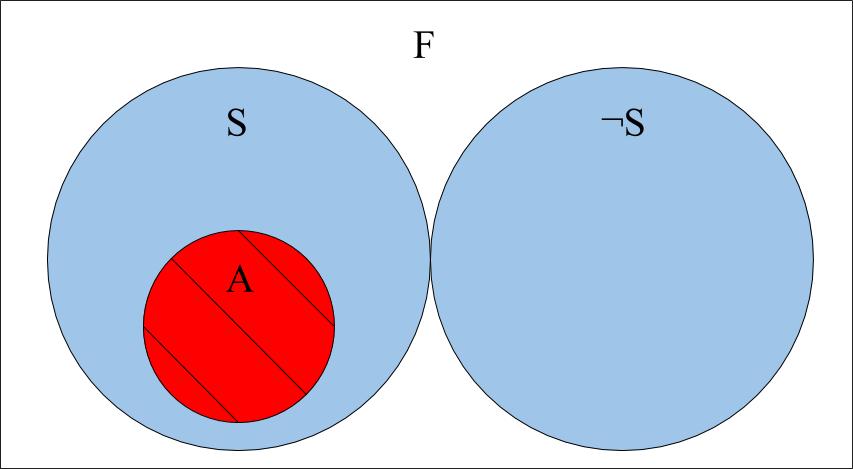}
    \caption{Classes of traffic flows ($F$) considered by \name. Attack traffic is shaded in red with diagonal hatches ($A$). Benign Traffic is shaded in solid blue. ($(S-A)\cup \neg S$).}
    \label{fig:traffic_classes}
\end{figure}

In our model, flows ($F$) are denoted by a source and a destination address $(src, dst)$. We assume that the flows originate and terminate outside of $G$, and categorize them into four subsets. Flows can be Attack ($A$), Suspicious ($S$), Not Suspicious (also known as Trusted) ($\neg S$), and Benign ($B$). We use the terms: good, benign, and legitimate interchangeably when referring to non-attack traffic. However, $B$ traffic is not necessarily trusted traffic, $\neg S$. The four classes are related as follows (see Figure~\ref{fig:traffic_classes}). All flows are either Suspicious or Trusted; $S \cup \neg S = F \And S \cap \neg S = \emptyset $. Attack traffic is a subset of Suspicious traffic; $A \subseteq S$. Trusted traffic is a subset of Benign traffic; $\neg S \subseteq B$. Benign traffic is $(S-A)\cup\neg S$. % Figure~\ref{fig:traffic_classes} illustrates the different classes of traffic as a Venn-diagram.

\subsection{Direct Attacks}
\label{sec:direct_attack}
In a direct attack, a large number of attackers flood a victim with traffic such that the victim cannot respond to legitimate users. From an attacker's standpoint, simple techniques (e.g., reflection and amplification~\cite{Reflection_Amplification, czyz2014taming}) can increase the strength of direct attacks without requiring additional resources. 

{\bf Scrubbing-based defense.}
In a traditional, scrubbing-based solution, anomalies in traffic patterns trigger an alarm when voluminous traffic that is bound for a targeted client enters the network. After the presence of the attack is known, the network reroutes all traffic bound for the target through a fixed set of hardware scrubbing appliances. %(see figure~\ref{fig:attack-model}). 
This rerouting introduces additional latency, and the bandwidth of the devices themselves adds a fixed limit to the throughput of traffic for legitimate senders exiting the scrubber. 

% \begin{figure}[htb!]
%     \centering
%     \includegraphics[width=0.5\columnwidth]{figures/Attack_Model.jpg}
%     \caption{Benign and attack traffic enters through boarder routers $a\_1$ though $a\_n$, and benign traffic is identified at router $x$. Router $x$ transmits benign traffic on a separate lambda from suspicious and attack traffic. At datacenter $d$ suspicious and benign traffic are scrubbed further, thus only benign traffic egresses to the client behind router, $v$.}
%     \label{fig:attack-model}
% \end{figure}

\begin{figure}[htb!]
    \centering
    \includegraphics[width=\columnwidth]{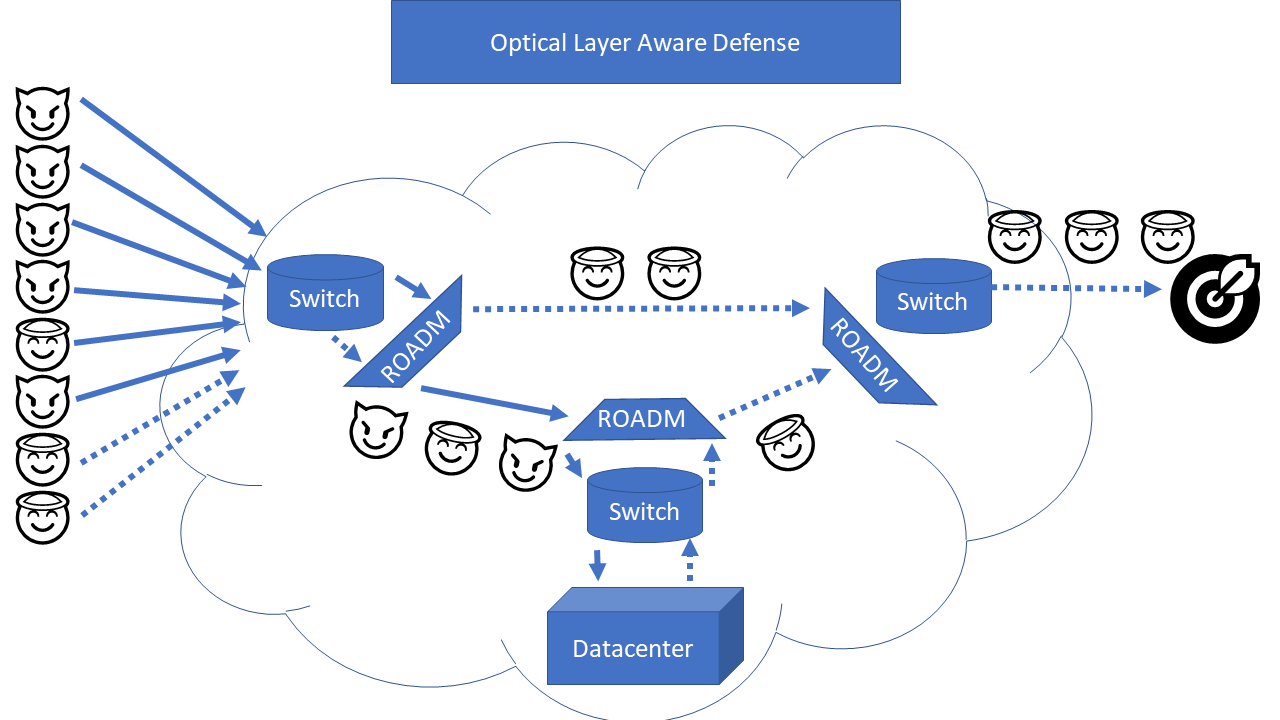}
    \caption{An \name identifies suspicious (solid) and non-suspicious (dashed) traffic. Then, physically separates both traffic types on distinct optical channels. It forwards trusted traffic directly to the destination, and suspicious traffic to the scrubbing datacenter.
    }
    \label{fig:controller}
    
\end{figure}

{\bf \name.} With \name, we can achieve physical isolation of traffic, diverting suspicious traffic $S$ through the scrubber and forwarding $\neg S$ directly to the client, as shown in Figure~\ref{fig:controller}. We can achieve this by switching $\neg S$ traffic to an alternate lambda {\em before} it can enter the datacenter. Then, a ROADM can be triggered to route the $\neg S$ wavelength directly to the destination. Now datacenter only scrubs suspicious traffic. 

% \begin{figure}[htb!]
%     \centering
%     \includegraphics[width=.5\columnwidth]{figures/Insider_datacenter.jpg}
%     \caption{There is a scrubber $s$ and egress router $e$ inside the datacenter $d$. An \name identifies benign traffic, giving that traffic a fast-pass around $s$. Scrubbed traffic joins $\neg S$ at $e$ and traverses the egress link to the client.}
%     \label{fig:attack-model2}
% \end{figure}

\subsection{Indirect Attacks}
\label{sec:indirect_attack}
% Indirect link-flooding attacks, such as the Crossfire attack~\cite{kang2013crossfire}, have grown in popularity for researchers and attackers over recent years. 
In an indirect attack (e.g. Crossfire attack~\cite{kang2013crossfire}), a coordinated group of attackers sends traffic to each other such that their communications over the Internet target a specific backbone link. Such attacks also minimize the throughput for traffic to the intended target by choosing a critical link and creating an abnormally high demand for that link. 

{\bf Spiffy-based defense.} Spiffy~\cite{kang2016spiffy} identifies indirect attackers by a bandwidth scaling operation on congested links. It utilizes an SDN controller and an optimization framework to maximize the bandwidth scaling ratio for all links in an ISP network. The particular topology of the network limits this ratio. With capacity reserved, Spiffy reroutes a fraction of traffic from the link under attack via an alternate path. Then, it monitors the rate-change behavior of flows on the alternate path to detect malicious senders.

{\bf \name.} In \name, we propose leveraging idle and reconfigurable (low-priority) wavelengths. This additional capacity can increase throughput for legitimate senders, and enable us to identify malicious flows more quickly. Furthermore, \name can decrease latency for trusted flows traversing the link under attack. The latency decreases because we establish the backup wavelength point-to-point for the link, rather than rerouting the suspicious flows through the network. When the enterprise detects a link-flooding attack (e.g., by~\cite{xue2014towards}) it should allocate all suspicious traffic ($\neg S$) to one wavelength, and the trusted traffic ($S$) through a dynamically provisioned backup wavelength. Then, the enterprise can apply the rate-increase monitoring from~\cite{kang2016spiffy} on the affected traffic to black-hole traffic from malicious senders while increasing capacity for trusted senders. In \S~\ref{sec:indirect_attack_throughput}, we analyze the throughput for legitimate senders over time with \name and with Spiffy, and in \S~\ref{sec:latency} we model the expected latency with Spiffy and compare it with an \name solution.

%We use the technique described in~\cite{kang2016spiffy}, identifying bots by their rate-change behavior given additional bandwidth. However, we differ from them Spiffy by using additional lambdas on the flooded point-to-point link, rather than rerouting a subset of the impacted flows through a sub-optimal path through the network.

\section{Early Promise of \name}
\label{sec:analysis}
In this section, we present the throughput and latency gains for legitimate senders under a variety of attack scenarios and with different mitigation strategies.

\begin{figure*}[htbp!]
    \center
    \includegraphics[width=\textwidth]{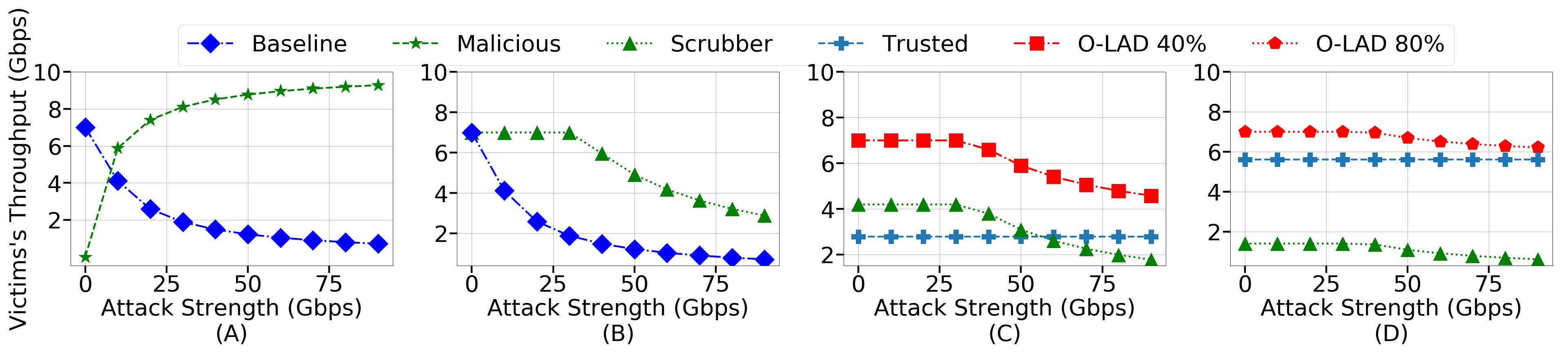}
    \caption{(A) Baseline Throughput for Legitimate Senders and Malicious attackers when no defense is deployed. (B) Throughput for Legitimate Senders when using a 40 Gbps scrubber, with a baseline for reference of improvement. (C) Throughput for legitimate senders when 40\% of good traffic is trusted. The \name line is the sum of the trusted line and scrubber line below. (D) \name's throughput when 80\% of traffic is trusted.}
    \label{fig:direct_attack_compairison}
\end{figure*}

\subsection{Throughput Gains For Direct Attacks}
\label{sec:througput_analysis}
We model the throughput for a voluminous direct attack under (i) a baseline scenario where no defense mechanism is implemented, (ii) when scrubbing is used, and (iii) when scrubbing is used with \name, as described in \S~\ref{sec:direct_attack}. We show that \name increases throughput for legitimate flows vs. (i) and (ii) during DDoS attacks of varying strength.

%\subsubsection{Baseline: Direct Attack with No Defense}

{\bf Baseline.} We analyze the throughput for benign and malicious traffic during a DDoS attack on an enterprise network. We assume that the victim has a 10 Gbps bandwidth connection. Furthermore, we suppose that when traffic demand to the target network meets or exceeds the allowed capacity, all traffic is prioritized equally (the target is incapable of distinguishing benign or malicious traffic). Finally, we assume that traffic to the victim has a historical trend of 70\% utilization. Thus 7 Gbps of traffic is from legitimate senders. Figure~\ref{fig:direct_attack_compairison} (A) shows that as the volume of the attack increases, the throughput for real users (goodput), quickly decreases. The goodput begins to fall as soon as the attacker sends enough traffic to saturate the network's bandwidth.

% \begin{figure}[hbp!]
%     \center
%     \includegraphics[width=0.98\columnwidth]{figures/Direct_Attack_No_Defense_10_7_100.png}
%     \caption{Baseline DDoS Direct attack scenario. When the total volume of traffic exceeds the bandwidth limitation of the victim, the attack successfully degrades goodput of the network.}
%     \label{fig:no_defense_direct}
% \end{figure}

%\subsubsection{State of the Art: Direct Attack with Scrubbing Defense}

{\bf Scrubbing.} The state-of-the-art solution for defending against direct attacks is to forward all flows to the victim through scrubbing appliances, either on-site at the victim's location, or within a carrier network. We model the target's traffic as it is re-routed through a scrubbing service (e.g.,~\cite{Scrubbing_CenturyLink}). Recall, the victim has 10 Gbps bandwidth and 7 Gbps of traffic is from legitimate users. Suppose scrubbing service has a fixed capacity of 40 Gbps. We assume that the scrubber is 100\% effective in removing malicious traffic. However, after the attack volume exceeds the scrubber's bandwidth, the total throughput for the victim decreases, just as it did in the baseline scenario. Figure~\ref{fig:direct_attack_compairison} (B) illustrates the limitation of the scrubbing defense mechanism. After the attack strength reaches 40 Gbps, services for the legitimate senders degrades. Note that attack strength required to degrade services increases over the baseline scenario. If the attacker's goal were to reduce the rate for legitimate sends to 3 Gbps, then the attacker would need to scale the attack from 20 Gbps to 80 Gbps---a 4x increase in cost for the attacker. 

% \begin{figure}[htbp!]
%     \center
%     \includegraphics[width=0.98\columnwidth]{figures/Direct_Attack_Scrubber_Defense_10_40_7_100.png}
%     \caption{The state of the art solution, scrubbing, stymies the attack until the attack volume exceeds the bandwidth limitation of the scrubber. Note: Goodput and Total are equal, as the scrubber is 100\% effective at filtering malicious traffic. Goodput decreases due to queuing delay at the scrubber.}
%     \label{fig:scrubber_defense_direct}
% \end{figure}

%\subsubsection{\name defense for Direct Attacks}

{\bf Scrubbing + \name.} Now, consider the throughput that the network can forward to the client with \name. 
% Let $T$ Gbps of ingress traffic is entering the system. Let $\hat{T}$ be the volume of traffic that we forward to the client. If we scrub 100\% of the ingress traffic, then the rate of traffic that we can forward to the client is limited by the bandwidth of the scrubbing appliances at the datacenter, $T^{D}$. However, if we can leverage reconfigurable lambdas to send trusted traffic, $T^{\neg{S}}$ around the scrubbers, as in figure~\ref{fig:attack-model2}, then the total bandwidth to the client is augmented with the trusted traffic.
% \begin{equation}
%     \hat{T} = T^{D} + T^{\neg{S}}
% \end{equation}
We assume that a fixed proportion of legitimate traffic can be trusted, regardless of the attack strength. For example, if the historical utilization of the service under attack is 7 Gbps, then a fraction of the senders who make up 7 Gbps of demand are labeled as trusted. With trusted traffic prioritized, we can forward it to the victim without involving a scrubber. Simultaneously, we will deliver all other suspicious traffic to the scrubber. 

Figures~\ref{fig:direct_attack_compairison} (C) and (D) illustrate the throughput for different classes of traffic (trusted, and suspicious, malicious, and total) when 40\% or 80\% of benign traffic is from trusted sources. We can see from the figure that goodput of the network asymptotically approaches the volume of trusted traffic as the strength of the attack grows. We also notice that the throughput for data leaving the scrubber approaches zero as the strength of the attack increases. The goodput of the network is the sum of trusted traffic and the scrubbers throughput. We argue that physically separating traffic on distinct wavelengths, and only sending suspicious traffic to the scrubber, increases the quality of service for the victim's network during a DDoS attack. 

% \begin{figure}[htbp!]
%     \center
%     \includegraphics[width=0.98\columnwidth]{figures/Direct_Attack_Fastpass_Defense_10_40_7_100_80.png}
%     \caption{\name DDoS defense for Direct attack. Assuming 80\% of benign traffic can be trusted before scrubbing. As the strength of the attack increases, the goodput of the network approaches the throughput of trusted traffic. The throughput of data exiting the scrubber approaches zero as the strength of the attack increases.}
%     \label{fig:FastPass_defense_80}
% \end{figure}

%\subsubsection{Throughput Comparison}

{\bf Improvements via \name.} Considering the three mitigation strategies, we see that \name pushes the boundary further for the strength of DDoS attacks that a network can tolerate. In Figure~\ref{fig:direct_attack_compairison} we see that when we trust 40\% and 80\% of the legitimate traffic, the impact of an attack is reduced significantly against scrubbing and the baseline. Specifically, consider a 40 Gbps DDoS attack. In the baseline scenario, throughput for legitimate traffic fell from 7 to $\sim$1.5 Gbps. With the scrubber, throughput fell to 6 Gbps. (Note: that the scrubber stopped being completely capable of defending the target when the attack strength was 33 Gbps, as the total of malicious and legitimate traffic was 40 Gbps). When the attack strength is 40 Gbps, the total traffic traveling through the scrubber is 47 Gbps. Therefore performance for legitimate users begins to be impacted. However, if 40\% of the 7 Gbps of good traffic never traverses the scrubber, then the load of the scrubber is reduced by 2.8 Gbps, and the performance for good traffic under this 40 Gbps attack only falls to $\sim$6.5 Gbps instead of 6 Gbps. If we route 80\% of good traffic (5.6 Gbps) around the scrubber, then a 40 Gbps attack has almost no effect on legitimate users. 

As the scale of the attack increases, so does the benefit in providing an optical layer-aware defense. When the attack strength reaches 100 Gbps in the baseline scenario, the throughput for legitimate senders falls to 0.65 Gbps, $\sim$9\% of its original strength. The scrubber alone, helped to keep throughput up to 2.6 Gbps, or $\sim$37\% of the full strength. If we forward 40\% of the legitimate traffic as trusted, the aggregate throughput increases from 2.6 Gbps to 4.4 Gbps, or 62\% 7 Gbps---an improvement of 25\% percent. Finally, if the network can identify 80\% of traffic as trusted, then throughput for legitimate senders is $\sim$6.2 Gbps, or 88\%---an improvement of 51\%. These early results show that \name can help increase throughput from  25\% to 51\% over the scrubber in this scenario. 

% \begin{figure}[htbp!]
%     \center
%     \includegraphics[width=0.98\columnwidth]{figures/Direct_attack_compairisons.png}
%     \caption{Throughput for legitimate traffic for baseline, scrubbing, and \name, where \name can identify 40\% and 80\% of benign traffic without scrubbing.}
%     \label{fig:direct_attack_compairison}
% \end{figure}

\subsection{Throughput Gains for Link-Flooding Attacks}
\label{sec:indirect_attack_throughput} 
Next, we consider the throughput of our network under attack from an indirect, link-flooding attack. Suppose that the target link capacity is 10 Gbps and senders on this link are guaranteed a 100 Mbps data rate. Further, suppose the attacker wants to reduce the rate for legitimate senders by ten-fold. Thus the attacker needs to generate 100 Gbps of attack traffic. According to the optimal attack strategy in~\cite{kang2016spiffy}, this attacker requires 10 thousand attack flows, sending data at 10 Mbps each, to reduce the capacity of the link to the intended level.

The response time for a network to mitigate the link-flooding attack with Spiffy is a factor of how much extra capacity can be reserved for fighting link flooding attacks in the layer-3 topology, $M_{network}$~\cite{kang2016spiffy}. Kang et al., show that 5 to 10 operations of {\em Temporary bandwidth expansion} (TBE) are required to identify 90\% of attack flows (based on experiments with ISP topologies; see Figure 15 in ~\cite{kang2016spiffy} for more details). Also, each of these operations takes $\sim$5 seconds. Therefore, we expect that the target link is to be congested for $\sim$25 to 50 seconds.

Using \name, the introduction of backup wavelengths to the network or acquisition of low-priority wavelengths for defenses could potentially increase $M_{network}$ to $M_{ideal}$, which is the volume of backup capacity required to identify all bots with one operation of TBE. If these backup wavelengths are added in the same order of time as a TBE operation, it is possible to mitigate the attack in $\sim$5 seconds. We think this is entirely possible, given the innovations in the optical community including programmable transceivers~\cite{history}, optical amplifiers for reconfigurable networks~\cite{ikhsan2018performance, xiang2018joint}, throughput gains of elastic transceivers in long-haul networks~\cite{ives2017throughput},  %The optics community is keenly interested in producing new solutions for stronger, more reconfigurable networks~\cite{strasser2010wavelength}. 
and recent momentum in the networking community~\cite{porter2013integrating, von2015coronet, durairajan2018greyfiber}. Thus, we implore researchers in the security community to consider optical layer when designing and modeling the next generation of solutions for DDoS attacks. 

\subsection{Latency Models and Improvements}
\label{sec:latency}

\begin{figure*}[htbp!]
    \center
    \includegraphics[width=1.0\textwidth]{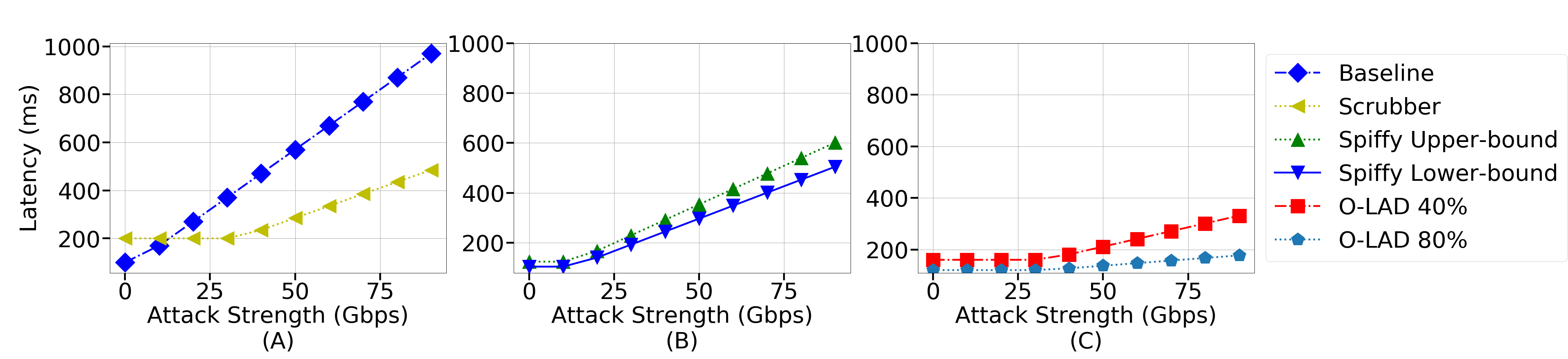}
    \caption{(A) Latency for legitimate traffic for baseline and scrubbing. (B) Latency for legitimate senders with Spiffy, where alternate path have an upper-bound additional latency of 24\%, and lower bound paths have 4\% additional latency. (C) \name, where 40\% and 80\% of legitimate users can be identified and routed around congestion. }
    \label{fig:latency_analysis}
\end{figure*}

Here, we describe the latency improvements with \name, first describing the baseline latency expectations formally for (i) no defense, (ii) scrubber-based defense for direct attacks, and (iii) Spiffy-based defense for indirect attacks. Then we describe a solution using \name. To start, we consider three classes of traffic and reason about the expected latency for each of them with \name for direct DDoS attacks: Trusted Traffic, ($\neg S$), suspicious traffic from legitimate senders ($S-A$), and their union, all good traffic ($(S-A)\cup \neg S$). 

{\bf Baseline.} For our baseline analysis of latency, we consider the metric as the product of the ratio of demand to capacity and expected a delay when there is no congestion.
% Similarly, the reduction of load at the scrubber is $T^{\neg S}$ Gbps.  We will vary the proportion of traffic that can be labeled $\neg S$ from 0 to 100 percent. The average latency, $\hat{L}$, for flows to the client is a weighted average of the latency for the proportion of traffic that travels the scrubber, and the ratio of those that do not.
% \begin{equation}
%     \hat{L} = \frac{(\delta) T^{\neg S} + (\delta+\epsilon)T^{S}}{T}
% \end{equation}
% Where $\delta$ is the baseline delay through the network from ingress to $v$, and $\epsilon$ is the additional latency incurred at the scrubber. 
\begin{equation}
    L_{Baseline} = 
\begin{cases}
        \delta        & \text{if $T \leq T^c$} \\
        \delta * T/T^c  & \text{ if$T > T^c$}
\end{cases}
\label{eq:latency_baseline}
\end{equation}

In equation~\ref{eq:latency_baseline}, $\delta$ is the baseline (propagation) delay through the network from ingress to the victim, $T$ is the aggregate demand for all flows $F$, and $T^c$ is the physical capacity of the congested path or victim. We show the baseline model, with the fixed capacity, $T^c$, of 10 Gbps, in Figure~\ref{fig:latency_analysis} (A). 

{\bf Scrubber.} To model latency for a scrubber, we introduce a new additive term, $\epsilon$, representing the additional latency incurred by traffic through the scrubber in Equation~\ref{eq:scrubber_latency}. We also replace the capacity of the client, $T^c$, with the capacity of the datacenter housing the scrubber, $T^D$. Figure~\ref{fig:latency_analysis} (A) shows the expected latency when a scrubber is used, assuming that the baseline latency ($\delta$) is 100 ms, the scrubber-induced delay ($\epsilon$) is 100 ms, and the capacity of the datacenter ($T^D$) is 40 Gbps. The slope of the latency curve is determined by the capacity of the network and the baseline latency ($d$). Precisely, it is the derivative of $L$ with respect to $T$, or $( 3\delta+\epsilon) / T^D$. Thus, a higher capacity implies a lower impact on latency. In this model, the scrubber can withstand up to $T^D$ Gbps of throughput before performance degrades. This protection comes at the cost of additional latency for re-routing all flows to the scrubber. Hence, the Scrubber solution's latency starts at 200 ms instead of the baseline 100 ms.
\begin{equation}
    L_{Scrubber} = 
\begin{cases}
    (\delta + \epsilon) & \text{if $T \leq T^D$} \\
    (\delta + \epsilon) (T/T^D)  & \text{if $T > T^D$}
\end{cases}
\label{eq:scrubber_latency}
\end{equation}

% \begin{figure}[htbp!]
%     \center
%     \includegraphics[width=0.5\columnwidth]{figures/Latency_Scrubber_Defense_10_7_100_40.png}
%     \caption{Latency for legitimate traffic for baseline and scrubbing.}
%     \label{fig:baseline_latency}
% \end{figure}

{\bf Spiffy.} To model latency for Spiffy, we recognize two critical factors described in~\cite{kang2016spiffy}. (1) Recall that Spiffy reduces congestion by expanding bandwidth with reserved supplies in the network. This factor, $M_{Network}$, is topology dependent. When evaluated on several real-world topologies, the potential bandwidth scaling factor, $M_{Network}$ (or $M_N$) was approximately two times the initial capacity (See figure 14 in~\cite{kang2016spiffy}). Therefore, after twice the initial capacity is exceeded, latency increases. (2) Spiffy uses alternate paths to forward traffic and identify attackers. We will refer this factor as $APL$, the percentage increase in the alternate path to the normal path. By~\cite{kang2016spiffy}, $APL$ is expected to be 4 to 24\% longer, so we say $APL$ is 0.04 to 0.24. We augment the scrubber's latency model for Spiffy by substituting $\epsilon$ with $\delta * APL$ as shown below.
\begin{equation}
    \epsilon = \delta * APL
    \implies
    \delta + \epsilon = \delta + \delta * APL \\
     = \delta * (1 + APL)
\end{equation}

We then substitute $T^D$ with $M_{Network}$ to obtain
\begin{equation}
    L_{Spiffy} = 
\begin{cases}
    (\delta * (1 + APL)) & \text{if $T \leq M_N$} \\
    (\delta * (1 + APL)) (T/M_{N})  & \text{if $T > M_{N}$}
\end{cases}
\label{eq:latency}
\end{equation}

Figure~\ref{fig:latency_analysis} (B) shows the expected latency for Spiffy~\cite{kang2016spiffy} during a link-flooding attack. Latency during low strength attacks (0 to 14 Gbps) is 104 ms to 124 ms, which is relatively close to the baseline (100 ms). This initial latency is better than the scrubber's, which started at 200 ms. After the attack traffic exceeds the reserved bandwidth, $M_N$, latency increases. 
% \begin{figure}[htbp!]
%     \center
%     \includegraphics[width=0.5\columnwidth]{figures/Latency_Spiffy_Defense_10_7_100.png}
%     \caption{Latency for legitimate traffic for baseline and scrubbing.}
%     \label{fig:baseline_latency_spiffy}
% \end{figure}

% \subsection{Latency Improvement with \name} 
% \label{sec:latency_hyperion}

{\bf \name.} The benefit of \name is its ability to separate ${\neg S}$ and $S$ flows, and use the physical separation to route $\neg S$ around congestion points (scrubbers or flooded links). To model the latency improvement with \name, we measure the weighted average of latency for all good traffic ($\neg S$ and $S-A$). Equation~\ref{eq:hyperion_latency} models this latency, using constructions from \S~\ref{sec:latency}. We present \name's latency as $L_{\name}^*$ where * is either $D$ for direct attacks or $I$ for indirect attacks. In the case of $L_{\name}^D$ we replace $L_{scrubber}$ with the relevant latency measure, $L_{Spiffy}$.
\begin{equation}
    L_{\name}^D = \frac{(L_{Baseline}) T^{\neg S} + (L_{Scrubber})T^{S-A}}{T^{\neg S}+T^{S-A}}
    \label{eq:hyperion_latency}
\end{equation}

The intuition is that trusted flows ($\neg S$) will have the baseline latency, and non-attack suspicious flows ($S-A$) will have a latency of the defense mechanism, either Spiffy or Scrubbing. The aggregate measure for latency for \name is the average of these two values, weighted by the proportion of traffic in each category.

Figure~\ref{fig:latency_analysis} (C) shows the latency values for varying attack strengths when 40 and 80\% of good traffic can be identified as trusted. We see that for a 100 Gbps attack, if 40\% of the good traffic is trusted, then latency drops to from 535 ms to 361 ms against the scrubber---a 33\% decrease. If 80\% of the good traffic can be trusted, then it falls to 187 ms---a 65\% decrease. These early results show promise for an optics-based solution for RAC in the face of DDoS attacks.

% Our \name solution reduces congestion without sending traffic on alternate layer-3 paths. Suspicious traffic in our model travels an alternate lambda between the same source and destination on the target link, and thus reduces congestion of the affected link without incurring any additional latency. 

% \input{sections/summary.tex}
%%%\input{sections/lab-based_experiments.tex}
\section{Future Outlook}

% {\bf TODO: Expand with details.}
An approach like \name opens up a number of interesting problems at the intersection of optical, security, and networking communities, which we outline below.

{\bf On the Feasibility of \name for Diverse DDoS Attacks.} Apart from the direct vs. indirect DDoS, grand challenges lie ahead in modeling and evaluating the gains of \name for combating other types of DDoS attacks. In particular, the feasibility of \name in defending (in)distinguishable, fixed vs. variable rate, volume-based, and protocol-conforming attacks calls for further research involving optical and security communities. 

{\bf Towards a Commercial, Industry-grade \name System.} In addition to understanding and evaluating the efficacy of \name via models, the lack of vendor-agnostic APIs (as discussed in \S~\ref{sec:challenges}) might impede the further development of \name into a full-fledged DDoS defense system. This calls for collaborations among optical, networking and security researchers and to create new partnerships e.g. between OpenConfig~\cite{openconfig}, enterprises, and security experts to solve the grand challenges in this front. Furthermore, the heterogeneity, scale, and dynamism of modern DDoS attacks require new testing frameworks and capabilities for \name to operate effectively against the growing DDoS landscape. 

{\bf \name for a Broader Class of Cyber Attacks.} While the goal of this paper is to make a case for optical layer-aware DDoS defense, we believe that the notion of optical layer awareness is beneficial for a broader class of cyber attacks. First, insider reconnaissance is an on-going problem since the topology can be mapped as shown by Achleitner et al.~\cite{achleitner2016cyber}. By keeping the performance and network objectives in mind, we believe \name can arbitrarily change wavelengths to effectively combat reconnaissance by providing cyber deception. In addition, complementary to NetHide~\cite{meier2018nethide}, we believe that \name can be used to combat targeted attacks by dynamically altering the underlying wavelengths and, hence, topologies.

\bibliographystyle{abbrv} 
\begin{small}
\bibliography{hotnets19}
\end{small}

\end{document}